\documentclass[12pt]{article}

\usepackage[paper=letterpaper,margin=1.5in]{geometry}

\usepackage{setspace}
\makeatletter
\AtBeginDocument{%
\let\oldsection\section%
\newcommand{\uppercasesection}[2][]{%
        \oldsection[#1]{\MakeUppercase{#2}}}%
\newcommand{\uppercasesectionstar}[1]{%
        \oldsection*{\MakeUppercase{#1}}}%
\def\section{%
        \@ifstar{\uppercasesectionstar}{\@dblarg{\uppercasesection}}}%
}
\makeatother

\usepackage[modulo,mathlines]{lineno}
\modulolinenumbers[1]
\firstlinenumber{1}
\linenumbers

\usepackage{paralist}

\usepackage{graphicx}
\usepackage{epstopdf}
\usepackage{epsfig}
\usepackage{psfrag}

\usepackage[%
            labelsep=newline,
            justification=RaggedRight,
            singlelinecheck=false,
            labelfont=bf,
            tableposition=top,
           ]{caption}
\makeatletter
\let\oldmakecaption\@makecaption
\makeatother

\usepackage[figoff]{figcaps}

\def\tablepagename{Tables}
\def\figurepagename{Figures}

\makeatletter
\def\phantomsection{\relax}
\let\oldfigurepage\@figurepage
\def\@figurepage{%
        \@ifundefined{tf@pof}{}{%
        \let\oldlabel\label%
        \let\oldinput\@input%
        \def\@input{\def\label{\oldlabel}\oldinput}%
        \phantomsection%
        \addcontentsline{toc}{section}{\figurepagename}}%
        \oldfigurepage%
        }
\let\@makefcaption\@makecaption
\let\oldtablepage\@tablepage
\def\@tablepage{%
        \@ifundefined{tf@pot}{}{%
        \clearpage%
        \phantomsection%
        \addcontentsline{toc}{section}{\tablepagename}}%
        \oldtablepage%
        }
\makeatother

\usepackage[subrefformat=subparens,labelformat=parens]{subfig}
\usepackage{amsmath,amssymb,amsfonts}

\makeatletter
\@ifpackageloaded{lineno}{%
\newcommand*\patchAmsMathEnvironmentForLineno[1]{%
  \expandafter\let\csname old#1\expandafter\endcsname%
        \csname #1\endcsname
  \expandafter\let\csname oldend#1\expandafter\endcsname%
        \csname end#1\endcsname
  \renewenvironment{#1}%
     {\linenomath\csname old#1\endcsname}%
     {\csname oldend#1\endcsname\endlinenomath}}%
\newcommand*\patchBothAmsMathEnvironmentsForLineno[1]{%
  \patchAmsMathEnvironmentForLineno{#1}%
  \patchAmsMathEnvironmentForLineno{#1*}}%
\AtBeginDocument{%
\patchBothAmsMathEnvironmentsForLineno{equation}%
}%
\@ifpackageloaded{amsmath}{%
\AtBeginDocument{%
\patchBothAmsMathEnvironmentsForLineno{align}%
\patchBothAmsMathEnvironmentsForLineno{flalign}%
\patchBothAmsMathEnvironmentsForLineno{alignat}%
\patchBothAmsMathEnvironmentsForLineno{gather}%
\patchBothAmsMathEnvironmentsForLineno{multline}%
}}{}%
}{}
\makeatother

\usepackage{natbib}
\usepackage{tocbibind}

\usepackage{varioref}
\labelformat{equation}{\textup{(#1)}}
\labelformat{enumi}{\textup{(#1)}}

\usepackage[%
            pdftitle={},
            pdfsubject={},
            pdfauthor={},
            pdfkeywords={},
            pdfstartview=FitH,
            bookmarks=true,bookmarksopen=true,
            breaklinks=true,
            colorlinks=false,
            pdfpagelabels=true,hypertexnames=true,
            plainpages=false,naturalnames=false,
            ]{hyperref}

\let\orgautoref\autoref
\providecommand{\Autoref}
        {\def\equationautorefname{Equation}%
         \def\figureautorefname{Figure}%
         \def\subfigureautorefname{Figure}%
         \def\Itemautorefname{Item}%
         \def\tableautorefname{Table}%
         \orgautoref}
 
\providecommand{\Autorefs}
        {\def\equationautorefname{Equations}%
         \def\figureautorefname{Figures}%
         \def\subfigureautorefname{Figures}%
         \def\Itemautorefname{Items}%
         \def\tableautorefname{Tables}%
         \orgautoref}
 
\renewcommand{\autoref}{\Autoref}
 
\providecommand{\autorefs}{\Autorefs}

\usepackage{mathptmx}
\usepackage{amssymb,amsfonts,amsmath}
\usepackage{color}
\usepackage{psfrag}
\usepackage{verbatim}
\usepackage{svn}
\usepackage[applemac]{inputenc}

\usepackage{bm}

\newcommand{\br}{\mathbf{r}}
\newcommand{\bv}{\mathbf{v}}

\begin{document}



\hypersetup{%
        pdftitle=
        {The title},
}
\title{Determining interaction rules in animal swarms}
\date{}
\maketitle

\vspace{-2 cm}
\noindent {\bf Anders Eriksson}

\noindent  {\small Department of Zoology, University of Cambridge, Downing St, Cambridge, CB2 3EJ, UK}

\noindent {\bf Martin Nilsson Jacobi, Johan Nystr\"{o}m, Kolbj{\o}rn Tunstr{\o}m } 

\noindent  {\small Complex Systems Group, Department of  Energy and Environment, Chalmers University of Technology, SE-41296 Gothenburg, Sweden}

\begin{abstract}
In this paper we introduce a method for determining local interaction rules in animal swarms. The method is based on the assumption that the behavior of individuals in a swarm can be treated as a set of  mechanistic rules. 

The principal idea behind the technique is to vary parameters that define a set of hypothetical interactions to minimize the deviation between the forces estimated from observed animal trajectories and the forces resulting from the assumed rule set. We demonstrate the method by reconstructing the interaction rules from the trajectories produced by a computer simulation.

\hypersetup{
        pdfkeywords=
        {swarming, behavioral rules, reverse engineering, force matching},
        }
\end{abstract}

\newpage



The collective motion of living organisms, as manifested by flocking birds, schooling fish, or swarming insects, presents a captivating phenomenon believed to emerge mainly from local interactions between individual group members. 
In part, the study of swarming and flocking aims to understand how animals use visual, audial and other cues to orient themselves with respect to the swarm of which they are part, and how the properties of the swarm as a whole depend on the behaviors of the individual animals. 
Also when addressing evolutionary questions of behaviour in swarms and flocks, such as the selective advantage of being bold or shy in response to a predator, it is important to  understand how the individuals behave based on the relation to their neighbours in the swarm or flock. 
For example, if the question is ``If the peripheral of the flock is more exposed to predators, do some individuals cheat the others by staying at the center of the flock where they are more protected?'', knowing the effective rules would make it easier to address questions regarding the evolutionary stability of the altruistic  behavior. 

Because flocks cannot be understood by studying individuals in isolation,
and are difficult to conduct controlled experiments on, understanding the behavioural patterns underlying flocking and swarming is especially challenging. Consequently, collective behavior has been extensively modeled particularly  using the agent-based modeling framework, where simple mechanistic behavioral rules are used to generate qualitatively realistic swarming behavior, 
 \citep[e.g][]{Aoki82, Reynolds87, Vicsek95, Parrish2,  Camazine,  Couzin02, Parrish,Viscido02B, Hoare04,Couzin, Cucker07, Mirabet07,  Romanczuk09, Yates}. The rules usually comprise three kinds of forces: A short-range force to avoid collisions with obstacles or other animals; a force adjusting the velocity to fit nearby individuals' velocities; and a force for avoiding being alone, e.g. by moving towards the average position of the nearby individuals. However, see e.g. \citet{Romanczuk09} for an alternative formulation. In addition, drag forces and noise are used to model the medium through which the individuals move, and external forces can be used to model interactions with terrain or predators.

The main strength of the agent-based modeling framework is the relative ease by which swarming behavior emerges from local interactions. This is however also its Achilles heel: Alternative mechanisms can generate visually similar swarming patterns. To reveal the effective interactions among swarming individuals of a specific species, several studies have introduced static quantitative observables, such as the distribution of inter-individual distances, swarm density, polarity, sharply defined edges and anisotropy, that can be used to compare the output of a model to observations of the biological system~\citep{Parrish,Viscido02A,Couzin03,Takagi04, Ballerini08B, Cavagna08, Huepe08,Sumpter08}.

These techniques can provide valuable insights into aspects of what type of interaction a group of animals use, but in general cannot be used to reveal both the type and the strength of the interactions. We suggest here how this can be made possible by reverse-engineering the interactions directly from observed trajectories. For this purpose, we adapt the force-matching (FM) technique originally introduced to obtain simplified force fields in complex molecular simulations~\citep{Ercolessi,Noid2008A,Noid2008B}. This technique minimizes the mean squared difference between observed total forces (estimated from the trajectories) and a set of force hypotheses building on knowledge about the system under consideration. For more details, see the "Materials and Methods" section.
%


The FM method relies on dynamical data, and to this date no such data of large swarms exists publicly available. While the technical challenges of obtaining trajectory data are demanding, there is currently an increased effort in collecting large scale data sets, as demonstrated by the STARFLAG project (see e.g.~\cite{Ballerini}), where flocks consisting of thousands of starlings above Rome were photographed, mapping the coordinates of the individual starlings. In the absence of field data, we use simulations of swarm models to produce underlying data. This way of testing the FM method also serves as a necessary proof of principle of the proposed framework. 


As a demonstration problem, we focus on distinguishing between two competing hypotheses for animal interactions \citep{Ballerini}: In the first alternative, called the `geometrical' hypothesis, swarm individuals base their movement decisions on the relative positions and velocities of neighboring individuals inside a sphere with fixed radius. The alternative hypothesis, called `topological', is that individuals use a fixed number of closest neighbors in the flock to perform the same task. The demonstration is motivated by the results in \citet{Ballerini}, where a different method was used to infer that the interactions among flocking starlings are topological. In that study it was also argued that  each bird is interacting with its $6$ or $7$ nearest neighbors, but the detailed nature of the interactions could not be inferred.
The ideal would be to use reconstructed trajectories of individual starlings, but this is unfortunately not available at present (Ballerini, personal communication). In lieu of this data we apply our method to simulated flocking dynamic under the geometrical and topological scenarios. 
Our analysis proceeds in two steps. First, we show that both the geometric and the topological scenario are very difficult to tell apart by fitting simulations of the geometric scenario using topological forces. Second, we demonstrate how including all forces in the fitting process solves this problem, and allows one to assess the relative power of either scenario to explain the varition in the forces observed in the sampled trajectories.


\section{Materials and Methods}

\subsection{Generation of trajectory data for force matching}


The trajectory data used in testing the FM methodology are generated from computer simulations of swarm models. Two different scenarios are set up and simulated: One in which each individual interact through geometric interactions, the other in which the individuals follow topological interaction rules. The geometrical scenario is modeled using the following equations of motion:
\begin{align}
		\frac{d^2 {\br}_i}{dt^2}  &=  \sum _{j \neq i} \Big[ f (r_{ij} ) \,\hat{\br} _{ij} + \alpha _1 \langle \bv_j - \bv_i \,|\, r_{ij} < r_c \rangle 
	 + \alpha _2 \langle \br_j - \br_i \,|\, r_{ij} < r_c \rangle \Big] -  \gamma \bv_i + \beta {\bm \zeta}_i (t).
	\label{eq:boids_standard}
\end{align}
Here ${\br}_i$ is the position of individual $i$, $r_{ij}$ is the distance between individuals $i$ and $j$, $\hat{{\br}}_{ij}$ is the normalized direction vector from $i$ to $j$, $f (r)$ is the collision avoidance force, $\alpha _1$ and $\alpha _2$ define the strength of the velocity matching respective positional preference forces, $\gamma$ is the drag force relative to the ambient medium and ${\bm \zeta} (t)$ is a noise vector. The geometrical scenario is manifested in the averages, e.g.  $\langle \br_j - \br_i \,|\, r_{ij} < r_c \rangle$, which involve all neighboring individuals within a cut-off radius $r_c$. The noise vector is necessary in combination with the drag force to set the average speed of the individuals. Each component of the noise vector is an independent random variable, uncorrelated between individuals and in time.

In the topological scenario, the equations of motion are nearly identical,
\begin{align}
	\frac{d^2 {\br}_i}{dt^2} &= \sum _{j \neq i} \Big[ f (r_{ij} ) \,\hat{\br} _{ij} + \alpha _3 \langle \bv_j - \bv_i \,|\, n_{ij} \leq  N \rangle 
	+ \alpha _4 \langle \br_j - \br_i \,|\, n_{ij} \leq N \rangle \Big] -  \gamma \bv_i + \beta {\bm \zeta}_i (t),
	\label{eq:boids_standard2}
\end{align}
but the averages now involve the $N$ closest neighbors.  $\alpha _3$ and $\alpha _4$, similar to $\alpha _1$ and $\alpha _2$, define the strength of the velocity matching respective positional preference forces. For simplicity, we take the collision avoidance force in both the geometrical and topological scenarios to be a linearly decreasing function:
\begin{equation}
f(r)= -\omega (1-r/R_\text{c})
\label{eq:linforce}
\end{equation}
when $r < R_{c}$, and is zero outside this range. The simulations are run with 200 individuals for $15 000$ unit time steps in a cube with side length $L=50$ and periodic boundary conditions. The parameters are set to: $\omega=0.1$, $R_\text{c}=5$, $\alpha_1 = \alpha_2 = \alpha_3 = \alpha_4 = 0.1$, $\gamma=0.1$, $\beta=0.1$, $r_\text{c}=4$, and $N=7$.

\subsection{The force matching method}
For a particle system where the trajectories are sampled with a time resolution adequate to decide the acceleration of each particle, the FM method is a useful tool to investigate the structure of the effective interactions. As stated in the introduction, the FM method is based on minimizing the mean squared difference between the observed force on a particle and the force resulting from a set of force hypotheses. Mathematically, by representing the observed total force on particle $i$ at time $t$ by the force vector $\mathbf{F}_{i}(t)$ and the corresponding force  predicted by hypothesis $h$ by $\tilde{\mathbf{F}}_{i}^{h}(t)$, this amounts to finding a minimum of the expression
\begin{equation}
	\left<  || \mathbf{F}_{i}(t) - \sum_{h}\tilde{\mathbf{F}}_{i}^{h}(t) ||^2 \right>,
\label{eq: force matching}
\end{equation}
where the average runs over local configurations in both space and time. The force at time $t$, $\mathbf{F}_{i}(t)$, is estimated using a finite difference approximation of the acceleration from three consecutive time steps: 
\begin{equation}
	 \mathbf{F}_{i}(t) = \frac{1}{\delta t^2}\big[\br_i(t+\delta t) - 2 \br_i(t) + \br_i(t-\delta t)\big]. 
	\label{eq:boids_obs}
\end{equation}

Setting up the force hypotheses generally requires knowledge about the system that is examined. Contrary to molecular particle systems, it is obviously impossible to explain the interactions between animals in terms of the fundamental theories of physics. This implies that the FM method applied to collective animal systems should focus not only on setting the parameters right for a given choice of interactions rules, but also to find and distinguish between biologically plausible force hypotheses. 

We show how the FM method can be used to assess the capability of competing mechanistic models to explain observed motion, by applying the method to the problem of distinguishing whether individuals in a swarm follows geometrical or topological interaction rules \citep{Ballerini}, and to provide an estimate of the actual interaction parameters.
First, suppose that repulsive interactions occur over a range $0 < r < R_c$. Partitioning this interval into $N_1$ equal bins, any smooth function can be approximated with a constant value within each bin. Thus, the estimated repulsive force can be written as  $\tilde{f}(r) = \sum_{k=1}^{N_1} a_{k} I_k(r)$, where $I_k(r)$ is an indicator function which is one when $r$ belongs to bin $k$ and is zero otherwise.
Second, to distinguish between the two scenarios for a simulated swarm, we let the hypothesized forces $\tilde{\mathbf{F}}_{il}^{h}$  include both the geometrical and topological scenarios. 
Using \autorefs{eq:boids_standard} and \ref{eq:boids_standard2} this leads to the following statistical  model for the force on particle $i$:
\begin{align}
		\tilde{\mathbf{F}}_{i}^{h} =& \sum_{j\neq i} \Big[ 
	\sum_{k=1}^{N_1} a_{k} \, I_k(r_{ij}) \,\hat{\br} _{ij} 
	 + \sum _ {k=1}^{N_2} b_{k} \,  \langle \br_j - \br_i \,|\, n_{ij} \leq  N_k \rangle 
		+ \sum _ {k=1}^{N_3} c_{k} \, \langle \br_j - \br_i \,|\, r_{ij} < r_{c,k} \rangle  \nonumber \\ 
	&+ \sum _ {k=1}^{N_2} d_{k} \,  \langle \bv_j - \bv_i \,|\, n_{ij} \leq  N_k \rangle
	+ \sum _ {k=1}^{N_3} e_{k} \, \langle \bv_j - \bv_i \,|\, r_{ij} < r_{c,k} \rangle  
		\Big] - \text{f} \, \bv_i ,
		\label{eq:boids_hyp}
\end{align}
where parameters $a_k$ to $f$ are unknown parameters to be estimated. They correspond to, respectively, collision avoidance ($a_k$), moving to average position using topological hypothesis ($b_k$), moving to average position using geometrical hypothesis ($c_k$), aligning velocity using topological hypothesis ($d_k$), aligning velocity using geometrical hypothesis ($e_k$), and the dissipative force ($f$). 
Because the stochastic forces are uncorrelated in time and between individuals, they affect only the variance and not the mean value in the minimization process, and need therefore not be included. This is of course true for the simulated swarm, but might not hold for real systems.

Note that multiple hypotheses are set up for both the geometrical and topological scenarios. As not only the strengths of the interactions are unknown, but also the interaction range, this motivates the inclusion of several hypotheses, spanning a wider interaction range. 
The cut-off values $r_{c,k}$ are $N_3$ values equally spaced between a minimum and maximum hypothetical cutoff radius (in this paper, $3$ and $5$, respectively). Note that the value of $r_c$ actually used in the simulations of the geometric scenario (i.e. $r_c = 4$) falls within this range, so that, if successful, the method is expected to find the correct parameter. In the topological scenario, we take $N_k = k$ for $k = 1, \ldots, 10$.
When applying the FM method to trajectory data from simulated swarms, with individuals following respectively \autoref{eq:boids_standard} (for the geometrical hypothesis) and \autoref{eq:boids_standard2} (for the topological hypothesis), we use the values $N_1 = N_2 = 10$, and $N_3 = 12$. We use the same value of $\delta t$ in the FM procedure as in the actual simulations. 

The minimization problem~\autoref{eq: force matching} can in general be solved iteratively, using e.g. the Newton-Raphson method. However, when $\tilde{\mathbf{F}}_{i}^{h}(t)$ is linear in its parameters, as is the case for \autoref{eq:boids_hyp}, the problem reduces to a linear least square problem $\min _x \| A x - y \|^2$, where $x$ is the parameter vector and $y$ is the observed accelerations. The minimization problem is equivalent to solving an overdetermined linear equation system $A x = y$ , where $A$ is an $n \times m$ matrix, $n \gg m$, and is solved by $x = (A ^T A )^{-1} A^T y$, see  \citep{pre96:num} for details. From the solution of the FM method, one can determine the relative importance of the hypothesized forces by the magnitude of the corresponding parameters,  so that the force that best correlates with the observed trajectories is favored over competing hypotheses. 



\section{Results}

\begin{figure}[h]
\center
\psfrag{x}[][]{Hypothesis}
\psfrag{y}[][]{Force coefficient}
\psfrag{a}[][]{a}
\psfrag{b}[][]{b}
\psfrag{c}[][]{c}
\psfrag{d}[][]{d}
\psfrag{e}[][]{e}
\psfrag{f}[][]{f}
\psfrag{forcefactor}[][]{Parameter value}
\begin{tabular}{l}
\raisebox{3cm}{\textbf{A}}
\includegraphics[width=129mm]{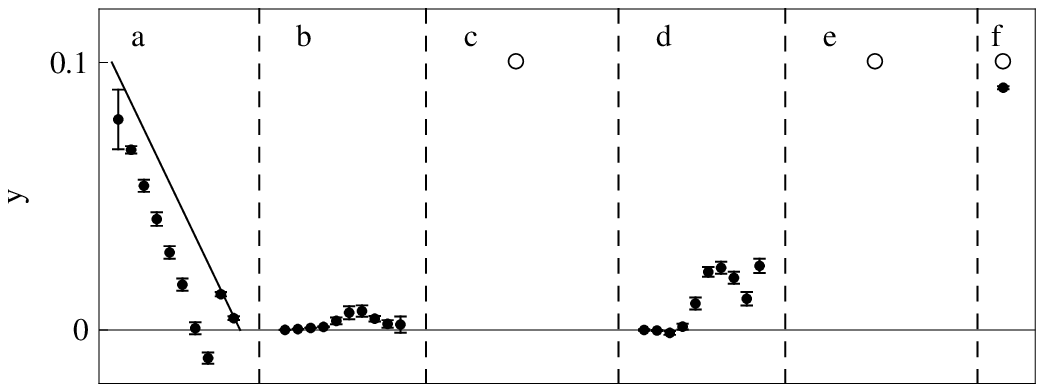} 
\hspace{-118mm}
\raisebox{3mm}{
{\small 0} \hspace{2mm} \, \raisebox{-2mm}{$r_{ij}$} \hspace{2mm}  \!{\small 5}\; \  
\,{\small 1}\!  \hspace{3mm} \!\raisebox{-2mm}{$N$}\!  \hspace{3mm} \!{\small 10}\,
\ \ \,{\small 3} \hspace{5mm}\raisebox{-2mm}{$r_c$} \hspace{4mm} {\small 5}
\ \ 
\,{\small 1}\!  \hspace{3mm} \!\raisebox{-2mm}{$N$}\!  \hspace{3mm} \!{\small 10}\,
\ \ \,{\small 3} \hspace{5mm} \raisebox{-2mm}{$r_c$} \hspace{4mm}  \!{\small 5}
}
\\[2mm]
\raisebox{3cm}{\textbf{B}}
\includegraphics[width=129mm]{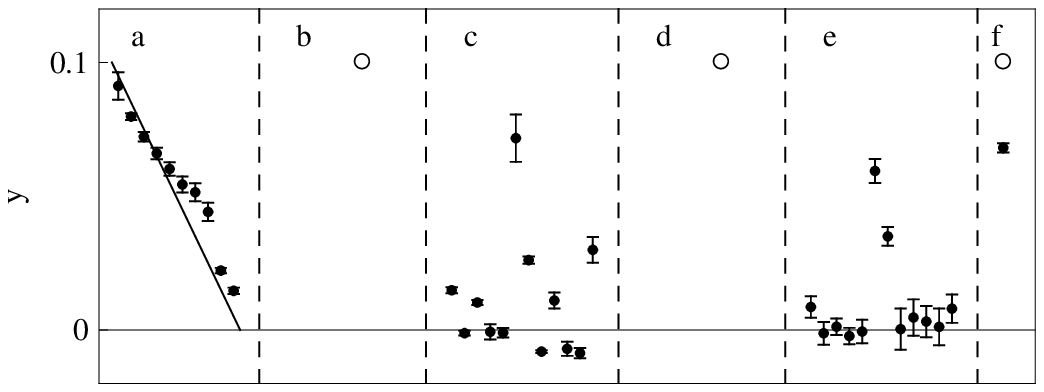}
\hspace{-118mm}
\raisebox{3mm}{
{\small 0} \hspace{2mm} \, \raisebox{-2mm}{$r_{ij}$} \hspace{2mm}  \!{\small 5}\; \  
\,{\small 1}\!  \hspace{3mm} \!\raisebox{-2mm}{$N$}\!  \hspace{3mm} \!{\small 10}\,
\ \ \,{\small 3} \hspace{5mm}\raisebox{-2mm}{$r_c$} \hspace{4mm} {\small 5}
\ \ 
\,{\small 1}\!  \hspace{3mm} \!\raisebox{-2mm}{$N$}\!  \hspace{3mm} \!{\small 10}\,
\ \ \,{\small 3} \hspace{5mm} \raisebox{-2mm}{$r_c$} \hspace{4mm}  \!{\small 5}
}
\end{tabular}
\caption{\label{fig:forces2}
}
\end{figure}

\begin{figure}[h]
\center
\psfrag{x}[][]{Hypothesis}
\psfrag{y}[][]{Force coefficient}
\psfrag{a}[][]{a}
\psfrag{b}[][]{b}
\psfrag{c}[][]{c}
\psfrag{d}[][]{d}
\psfrag{e}[][]{e}
\psfrag{f}[][]{f}
\psfrag{forcefactor}[][]{Parameter value}
\begin{tabular}{l}
\raisebox{3cm}{\textbf{A}}
\includegraphics[width=129mm]{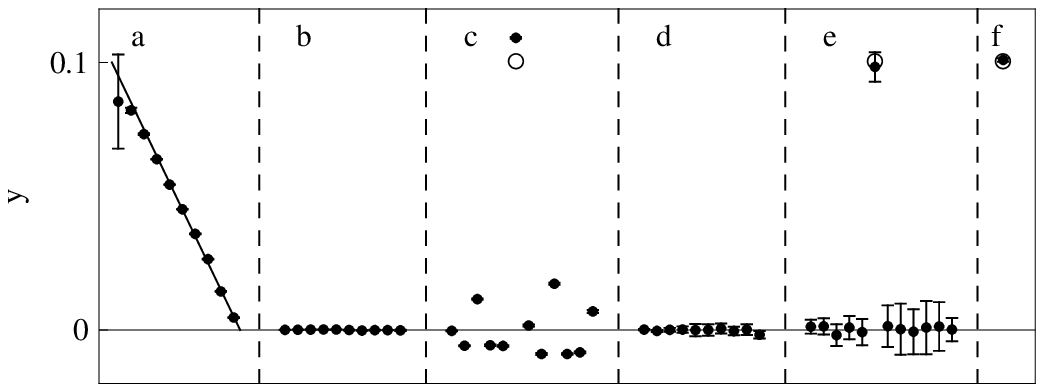} 
\hspace{-118mm}
\raisebox{3mm}{
{\small 0} \hspace{2mm} \, \raisebox{-2mm}{$r_{ij}$} \hspace{2mm}  \!{\small 5}\; \  
\,{\small 1}\!  \hspace{3mm} \!\raisebox{-2mm}{$N$}\!  \hspace{3mm} \!{\small 10}\,
\ \ \,{\small 3} \hspace{5mm}\raisebox{-2mm}{$r_c$} \hspace{4mm} {\small 5}
\ \ 
\,{\small 1}\!  \hspace{3mm} \!\raisebox{-2mm}{$N$}\!  \hspace{3mm} \!{\small 10}\,
\ \ \,{\small 3} \hspace{5mm} \raisebox{-2mm}{$r_c$} \hspace{4mm}  \!{\small 5}
}
\\[2mm]
\raisebox{3cm}{\textbf{B}}
\includegraphics[width=129mm]{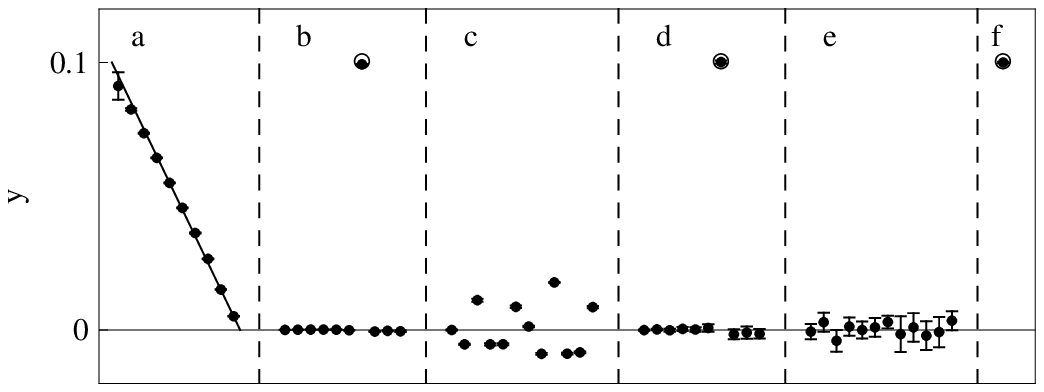}
\hspace{-118mm}
\raisebox{3mm}{
{\small 0} \hspace{2mm} \, \raisebox{-2mm}{$r_{ij}$} \hspace{2mm}  \!{\small 5}\; \  
\,{\small 1}\!  \hspace{3mm} \!\raisebox{-2mm}{$N$}\!  \hspace{3mm} \!{\small 10}\,
\ \ \,{\small 3} \hspace{5mm}\raisebox{-2mm}{$r_c$} \hspace{4mm} {\small 5}
\ \ 
\,{\small 1}\!  \hspace{3mm} \!\raisebox{-2mm}{$N$}\!  \hspace{3mm} \!{\small 10}\,
\ \ \,{\small 3} \hspace{5mm} \raisebox{-2mm}{$r_c$} \hspace{4mm}  \!{\small 5}
}
\end{tabular}
\caption{\label{fig:forces}
}
\end{figure}

Using trajectory data from simulations of swarming particles, following geometrical and topological interaction rules [\autorefs{eq:boids_standard} and \ref{eq:boids_standard2} in \emph{Materials and Methods}, respectively], we test the adapted FM method's performance for reconstructing known interactions.
The results are summarized in \Autorefs{fig:forces2} and \ref{fig:forces}. Black dots and intervals show averages and standard deviations, respectively, from eight independent simulations.
The figures are divided into sections for the different parameter groups, separated by dashed lines, showing the estimated coefficients of the different forces. From left to right in each panel: Section a shows the estimated collision avoidance force as a function of distance $r_{ij}$ between the animals.
Sections b and d shows the coffecients of the forces `moving to average position' and `velocity alignment' under the topological hypothesis, respectively, as a function of the maximum number of interacting neighbours ($N$).
Sections c and e are analogous to sections b and d, respectively, but shows the coefficients under the geometric hypothesis, as a function of the cutoff-radius $r_c$.
Finally, section f shows the strength of the dissipative force.
The forces corresponding to the true scenario simulated is shown as a solid line in section a, and by solid circles in panel a-f. The circles show the values of the coefficients used in the simulations (all other coefficients are zero). 

\Autoref{fig:forces2} shows how the method performs when simulating one scenario but fitting it to the other scenario.
In panel~$\textbf{A}$ the particles move according to the geometrical scenario, but only the topological forces are fitted. Conversely, in panel $\textbf{B}$ the particles move according to the topological scenario, but only the geometrical forces are fitted.
In both cases, the coefficients of the fitted forces are significantly different from zero, and if fitted independently one would therefore conclude that the wrong hypothesis is supported by the data. 

\Autoref{fig:forces} shows the same simulations as \Autoref{fig:forces2}, but with all coefficions fitted simultaneously. 
Under both the geometric and topological scenario, the method identifies the forces used in the simulations and correctly estimates their magnitude. Together, these results demonstrate the importance of fitting all parameters simultaneously. In general, if the scope of the range of forces tested against is too narrow, one may be mislead into accepting a false hypothesis because parameters can be found that generate a reasonable fit. However, when more than one hypothesis is included the magnitudes of the fitted forces show how much of the variation in the trajectories can be attributed to the hypothesis the forces correspond to: In \Autoref{fig:forces} it is apparent from the relative magnitude of the coefficients which hypothesis is favoured over the other.

Despite appearances the scattering of coefficients with low magnitude in section c of both panels in \Autoref{fig:forces} is not random; not only are the coefficients significantly different from zero but take very similar values in both panel A and panel B. This structure is an effect of the repulsive force being equal in both cases. 


\section{Discussion}
\label{sec:disc}



Our findings can be summarized in two main points: First, 
we suggest to adapt and apply the force matching  method from multi-scale molecular dynamics to test and calibrate hypotheses about the local rules governing collective motion in swarms and schools. Second, based on swarm simulations, we demonstrate that when using the FM method on observed data, it is essential to include all candidate hypotheses simultaneously in the regression. Because several different mechanisms often give rise to similar behavior, the candidate hypotheses can give speciously good results when fitted independently, and it is therefore difficult to say which is the more likely candidate. When fitted simultaneously, however, the strengths of the different forces provide a measure of the explanatory power of the particular hypothesis in relation to the competing hypotheses.

The force matching is efficient if all individuals can be classified, with individuals in a class following identical rules (in this paper we have only considered one class). In practice a classification could  be based on e.g. morphology, such as size. However, if the rule variation is caused by traits that are harder to distinguish, e.g. genetic variation, it poses a much larger challenge to the method. Another possible problem is that the interaction rules could be non-deterministic. However, this is not a fundamental problem, since it should be relatively straight forward to expand the framework to include stochastic interaction rules. This is currently work in progress. 

The need for high temporal resolution may seem as a severe restriction on the FM method. In practice however, a mitigating factor is that trajectories need only be piecewise continuous and must not necessarily include all individuals in the swarm. For each data point we need only three images closely spaced in time [c.f.~\autoref{eq:boids_obs}], from which accelerations and velocities can be calculated in addition to the distances between the animals. Thus, if the setup allows for collecting images in bursts, one will obtain piecewise estimates of the quantities necessary to apply the force matching. As long as the rules underlying the observed trajectories do not change, it does not matter that there are gaps in the observed trajectories -- although the data can be used more effectively if continuous trajectories should be available.

The method presented in this paper should be seen as a theoretical basis for how to infer interaction rules in animal swarms but it could also possibly find general applicability in a wide variety of systems often modeled using mechanistic rules. Examples other than swarming include human crowding~\citep{Helbing00}, traffic~\citep{Helbing98}, and possibly finance~\citep{LeBaron02}.  While the method has been shown to work for simulated swarms, it has not yet been applied to field data, and therefore it is currently not possible to conclude how well the method handles the natural variability found in biological systems. 
Collecting dynamical data is a technical challenge, and at present such data is not available (at least not publicly). Several efforts in this direction is however  underway and we are currently involved in testing the proposed method on trajectories of schooling fish.

\bibliography{articles}

\section{Figure Legends}
%

\noindent {\bf \autoref{fig:forces2}}\\
Force parameters for two models estimated from trajectories using the FM method with only parts of the force hypotheses in \autoref{eq:boids_hyp}. 
In $\textbf{A}$, the FM method using only topological hypothesis applied to data from simulation with geometrical interactions [\autoref{eq:boids_standard} with $r_c=4$]. In $\textbf{B}$, vice versa [topological interactions according to \autoref{eq:boids_standard2} with $N = 7$]. See the text for the remaining parameters. The results were obtained from averaging over 8 simulations run with $200$ individuals for $15 000$ time steps in a cube with side length $L=50$ and periodic boundary conditions. 
The different sections labelled a, b, c, d, e, and f correspond to the parameters $a_k$ to $f$ in \autoref{eq:boids_hyp}. Within each section, $k$ increases from left to right. Each letter represents a behavioral rule, and the subscripts corresponds to force hypotheses: collision avoidance (a), moving to average position using topological rule (b), moving to average position using geometrical rule (c), aligning velocity using topological rule (d), aligning velocity using geometrical rule (e), and dissipative force (f). The parameters estimated with the FM method are plotted as black markers ($\bullet$) with error bars ($\pm$ one standard deviation). In section a, the repulsive force used in the simulations, see \autoref{eq:linforce}, is plotted as a solid line. In the remaining sections, the exact parameter values used in the simulations are plotted as empty circles ($\circ$). Only the non-zero values 
are shown.\\ 

\noindent {\bf \autoref{fig:forces}}\\
Estimating all force parameters simultaneously, from simulations of the geometric scenario (panel A) and the topological scenario (panel B).
The results were obtained from averaging over 8 simulations run with $200$ individuals for $15 000$ time steps in a cube with side length $L=50$ and periodic boundary conditions.
See~\autoref{fig:forces} for explanation of symbols, an the text for parameter values.


\clearpage
\section{Figures}


\end{document}